\documentclass{article}
\usepackage{spconf,amsmath,amssymb,amssymb,graphicx,multirow,multicol,bm,kotex,comment,hyperref,booktabs}

\usepackage{tabularx} 

\title{BEAT2AASIST model with layer fusion for ESDD 2026 Challenge}

\name{\begin{tabular}[t]{c}
Sanghyeok Chung$^{1}$, Eujin Kim$^{2}$, Donggun Kim$^{1}$, Gaeun Heo$^{2}$, Jeongbin You$^{1}$, Nahyun Lee$^{3}$\\
Sunmook Choi$^{4}$, Soyul Han$^{5}$, Seungsang Oh$^{1}$, Il-Youp Kwak$^{2,3}$
\end{tabular}}
\address{$^{1}$ Department of Mathematics, Korea University, South Korea \\
$^{2}$ Department of Statistics and Data Science, Chung-Ang University, South Korea\\
$^{3}$ Department of Smart Cities, Chung-Ang University, South Korea\\
$^{4}$ Center for Applied Mathematics, Cornell University, USA\\
$^{5}$  Department of Big Data Application, Hannam University, South Korea}

\begin{document}
%
\maketitle
\begin{abstract}
Recent advances in audio generation have increased the risk of realistic environmental sound manipulation, motivating the {\em ESDD 2026 Challenge\/} as the first large-scale benchmark for Environmental Sound Deepfake Detection (ESDD). We propose {\em BEAT2AASIST\/} which extends BEATs-AASIST by splitting BEATs-derived representations along frequency or channel dimension and processing them with dual AASIST branches. To enrich feature representations, we incorporate top-$k$ transformer layer fusion using concatenation, CNN-gated, and SE-gated strategies. In addition, vocoder-based data augmentation is applied to improve robustness against unseen spoofing methods. Experimental results on the official test sets demonstrate that the proposed approach achieves competitive performance across the challenge tracks.
\end{abstract}

\begin{keywords}
ESDD, BEATs, Multi-layer fusion, Vocoder
\end{keywords}

\section{Introduction}\vspace{-2mm}

Recent advances in text-to-audio (TTA) and audio-to-audio (ATA) generative models have enabled highly realistic synthesis of speech and environmental sounds, benefiting applications such as VR/AR and media production, while also introducing new security risks through malicious sound manipulation. These concerns have increased the need for ESDD.

Most existing work has focused on speech-based deepfake detection, where self-supervised (SSL) foundation models such as wav2vec2~\cite{NEURIPS2020_92d1e1eb}, HuBERT~\cite{9585401}, and WavLM~\cite{9814838} serve as core front-end backbones. Combined with multi-layer fusion and vocoder-based data augmentation \cite{martindonas24_interspeech,wu24b_interspeech,10448453,10446331,10094779,han2023}, these models have achieved state-of-the-art performance in major challenges ASVspoof5~\cite{WANG2026101825} and ADD2023~\cite{9123b6b9056f4f0291698dc291d0fc85}.

ESDD is more challenging due to wide variability of acoustic conditions and diverse environmental sound categories, motivating the development of SSL foundation models for environmental audio such as BEATs~\cite{pmlr-v202-chen23ag} and EAT~\cite{10.24963/ijcai.2024/421}. This trend is reflected in the {\em ESDD 2026 Challenge\/}, which evaluates both generalization to unseen TTA/ATA generators (Track~1) and performance under extremely low-resource black-box settings (Track~2). The official BEATs-AASIST baseline integrates the BEATs encoder for feature extraction with the AASIST~\cite{jung2021aasistaudioantispoofingusing} graph-based anti-spoofing back-end. 

In this paper, we propose {\em BEAT2AASIST\/} building on BEATs encoder and incorporating AASIST with a dual-branch architecture to capture complementary frequency- and channel-specific patterns. Multi-layer fusion and vocoder-based augmentation further enhance feature discrimination and robustness against
diverse spoof types. Experiments on the EnvSDD dataset demonstrate strong performance on both tracks of
the ESDD 2026 Challenge.\vspace{-1mm}

\section{Methodology}
\label{sec:models}\vspace{-2mm}

\begin{figure}[t]
\centering
\centerline{\includegraphics[width=240pt]{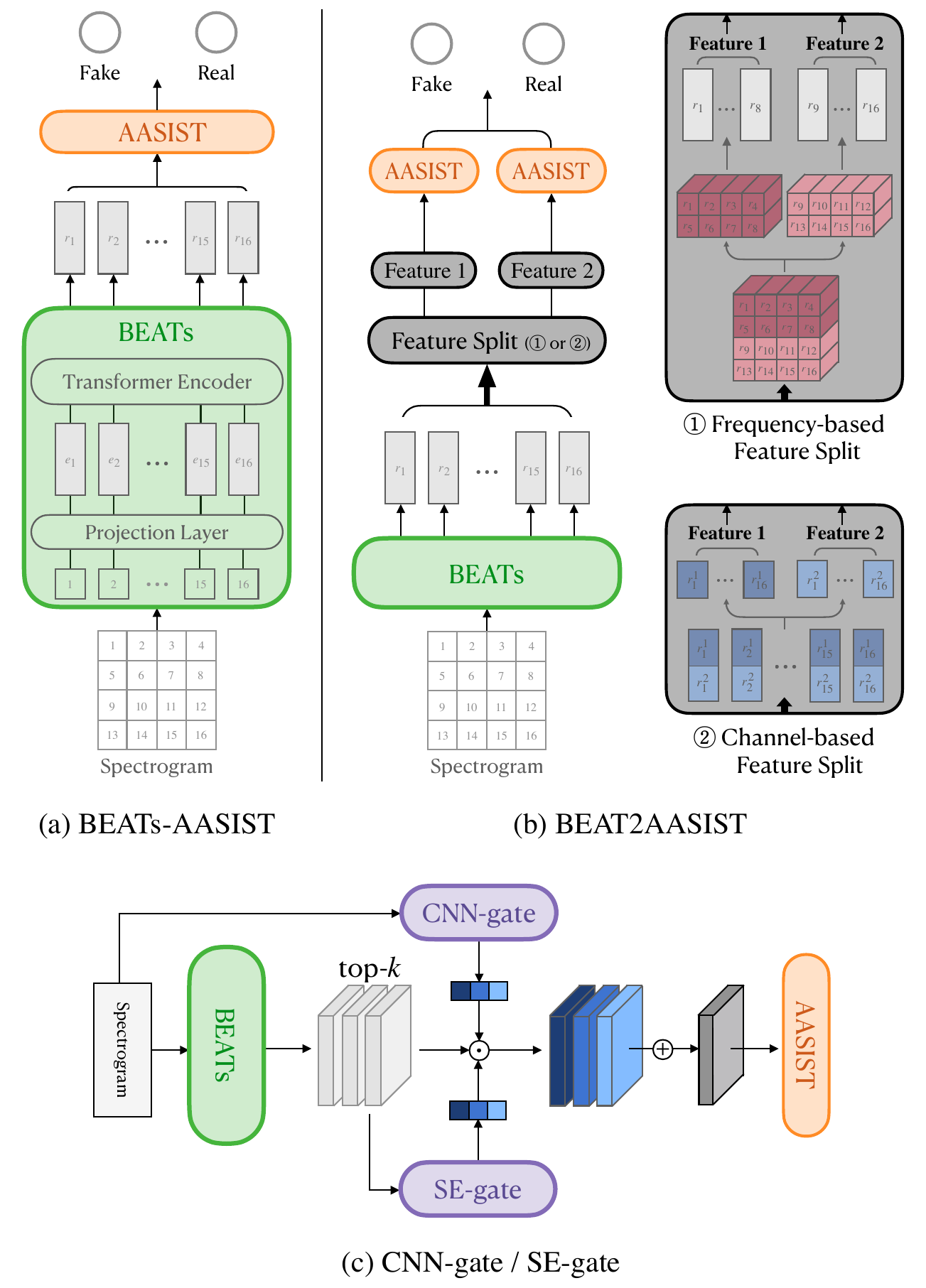}} 
\vskip -0.1in
\caption{BEAT2AASIST framework and fusion strategies}
\label{fig1}
\vskip -0.2in
\end{figure}

\subsection{BEAT2AASIST}\vspace{-1mm}

BEAT2AASIST extends BEATs-AASIST baseline (Fig.~\ref{fig1}(a)) by splitting BEATs-derived features and processing them through two independent AASIST branches (Fig.~\ref{fig1}(b)). BEATs receives a mel-spectrogram, divides it into $16\! \times\! 16$ size patches, and embeds each into a $768$-dimensional token, producing a feature sequence of shape ($B$, $H\! \times\! W$, $768$) with the batch size $B$ and the patch number $H$\,(or $W$) along the frequency\,(or time) axis. Two splitting strategies are used.\vspace{1mm}

\noindent \textbf{Frequency-based splitting.} 
The feature tensor is reshaped to ($B$, $H$, $W$, $768$) and split along the frequency axis into high- and low-frequency halves. Each part is reshaped back to ($B$, $H/2\! \times\! W$, $768$) and passed into its own AASIST network. The two outputs are concatenated to form the final prediction. This design enables the model to learn frequency-dependent spoofing cues and enhances sensitivity to spectral artifacts.\vspace{1mm}

\noindent \textbf{Channel-based splitting.} 
The feature tensor is split along the channel dimension into two parts of size ($B$, $H\! \times\! W$, $384$).
Each half is passed into an independent AASIST network, and the resulting outputs are concatenated to form the final decision. The model can capture channel-wise cues from BEATs.\vspace{-1mm}

\subsection{Multi-layer fusion}\vspace{-1mm}

Recent studies have shown that relying solely on the final transformer layer of large pre-trained models is often suboptimal, as fusing multiple layers generally yields richer and more robust representations. Consistent with this finding, the SUPERB benchmark~\cite{10502279} and WavLM adopt learnable layer-fusion mechanisms that assign adaptive weights across transformer layers. Motivated by these findings, we enhance the expressiveness of BEATs features by aggregating the top-$k$ layer representations via three fusion strategies (Fig.~\ref{fig1}(c)).\vspace{1mm}

\noindent \textbf{Concatenation fusion.} 
The top-$k$ layer outputs are concatenated along a new dimension into a tensor of shape ($B$, $k$, $H\! \times\! W$, $768$), preserving information from all selected layers.\vspace{1mm}

\noindent \textbf{CNN-gated fusion.} 
Fusion weights for the top-$k$ layers are computed from the input mel-spectrogram using a CNN-gate with three 2D convolution layers, followed by global average pooling and a linear\,+\,softmax layer, enabling input-dependent emphasis on more informative transformer layers.\vspace{1mm}

\noindent \textbf{SE-gated fusion.} 
Instead of using the mel-spectrogram, this method derives fusion weights directly from the top-$k$ outputs using an SE-gate. By modeling global channel-wise dependencies across $k$ layers, the fused representation captures more globally coherent acoustic cues.\vspace{-1mm}

\subsection{Vocoder augmentation.}\vspace{-1mm}

To improve robustness against diverse spoofing patterns, we apply vocoder-based augmentation. Prior studies by Yamagishi et al. have shown that neural vocoders such as HiFi-GAN~\cite{kong2020hifigangenerativeadversarialnetworks} can efficiently generate spoofed audio via copy-synthesis of bona fide samples without requiring on full TTS or voice conversion systems~\cite{10446331,10094779}. Following this approach, we synthesize spoofed audio using several high-quality, publicly available vocoders, including Hifi-GAN, BigV-GAN~\cite{lee2023bigvganuniversalneuralvocoder}, Univnet~\cite{jang2021univnetneuralvocodermultiresolution}. These GAN-based vocoders convert mel-spectrograms into waveform audio, producing diverse synthesis artifacts, enabling the model generalize better to unseen deepfake attacks and black-box conditions.\vspace{-1mm}

\section{Experiments}
\label{sec:experiments}\vspace{-2mm}

\noindent \textbf{Dataset.} 
We evaluate our system on the EnvSDD dataset~\cite{yin2025esdd2026environmentalsound}, which provides 45.25 hours of real audio and 316.7 hours of fake audio. Track 1 contains 27,811 real and 111,244 TTA/ATA-generated fake samples for training, with a test set of 1,000 real and 3,000 fake samples. Track 2 adds 270 real and 1,083 black-box fakes for training, and its test set includes 1,994 real and 7,980 black-box fakes.\vspace{1mm}

\begin{table}[t]
\centering
\footnotesize
\vskip -0.1in
\caption{BEAT2AASIST performance on Tracks 1 and 2.}
\label{table:1}
\begin{tabularx}{\linewidth}{
    p{0.2cm}
    >{\centering\arraybackslash}X
    >{\centering\arraybackslash}p{1.19cm}
    >{\centering\arraybackslash}p{1.15cm}
}
\toprule
\textbf{\#} & \textbf{Description for Track 1} & \textbf{Eval EER} & \textbf{Test EER} \\
\midrule
1 & \scriptsize BEATs-AASIST & 1.69\% & 1.88\% \\
2 & \scriptsize BEAT2AASIST, Freq., SE-gate ($k\!=\!4$) & 1.69\% & 1.92\% \\
3 & \scriptsize BEAT2AASIST, Channel, SE-gate ($k\!=\!10$) & 1.66\% & 1.70\% \\
\midrule
4 & \scriptsize Ensemble (0.4$\times$\#\,1 + 0.6$\times$\#\,3) & N/A & \textbf{1.60}\% \\
5 & \scriptsize Ensemble (0.3$\times$\#\,2 + 0.7$\times$\#\,3) & 1.66\% & 1.62\% \\
\bottomrule \toprule
\textbf{\#} & \textbf{Description for Track 2} & \textbf{Eval EER} & \textbf{Test EER} \\
\midrule
1 & \scriptsize BEATs-AASIST, concat. ($k\!=\!4$) / N/A & 0.61\% & 0.75\% \\
2 & \scriptsize BEAT2AASIST, Freq., CNN-gate ($k\!=\!4$) & 0.42\% & 0.46\% \\
3 & \scriptsize BEAT2AASIST, Freq., CNN-gate ($k\!=\!10$) & N/A & 0.60\% \\
\midrule
4 & \scriptsize Ensemble (0.4$\times$\#\,1 + 0.6$\times$\#\,2) & 0.30\% & 0.40\% \\
5 & \scriptsize Ensemble (0.2$\times$\#\,1 + 0.5$\times$\#\,2 + 0.3$\times$\#\,3) & N/A & \textbf{0.35}\% \\
\bottomrule
\end{tabularx}
\vskip -0.1in
\end{table}

\noindent \textbf{Experimental setup.}
All parameters of BEATs-iter3 pre-trained on AudioSet are fine-tuned within BEAT2AASIST. Audio clips are trimmed or repeated to 4 seconds and converted into mel-spectrograms with 25 ms frame length, 10 ms frame shift, and 128 mel bins. SpecAug~\cite{Park_2019} and vocoder-based augmentation are applied  (HiFi-GAN for Track 1; HiFi-GAN, BigV-GAN, UnivNet for Track 2). The batch size was 32, and training was
conducted for 20 epochs. Because the dataset contains more fake than real audio, class weights (fake, real)\,$=$\,(0.1, 0.9) to the cross-entropy loss are applied.\vspace{1mm}

\noindent \textbf{Experimental results.}
We report Equal Error Rate (EER) for both progress-phase (Eval EER) and final ranking-phase (Test EER). Table~\ref{table:1} summarize Tracks 1 and 2 performance, covering individual models (Systems \#\,1\,$\sim$\,3) and their ensemble variants. Our models demonstrate competitive performance across both tracks, achieving a third-place ranking on Track~2.

\section*{Acknowledgment}
This work was supported by the National Research Foundation of Korea(NRF) grant funded by the Korea government (MSIT) (No. RS-2024-00341075, RS-2023-00208284)

\bibliographystyle{IEEEbib}
\bibliography{refs}

@inproceedings{NEURIPS2020_92d1e1eb,
 author = {Baevski, Alexei and Zhou, Yuhao and Abdelrahman Mohamed et al.},
 booktitle = {in NeurIPS},
 pages = {12449--12460},
 title = {wav2vec 2.0: A Framework for Self-Supervised Learning of Speech Representations},
 url = {https://proceedings.neurips.cc/paper_files/paper/2020/file/92d1e1eb1cd6f9fba3227870bb6d7f07-Paper.pdf},
 volume = {33},
 year = {2020} }

@ARTICLE{9585401,
  author={Hsu, Wei-Ning and Bolte, Benjamin and Yao-Hung Hubert Tsai et al.},
  journal={IEEE/ACM TASLP}, 
  title={HuBERT: Self-Supervised Speech Representation Learning by Masked Prediction of Hidden Units}, 
  year={2021},
  volume={29},
  number={},
  pages={3451-3460},
  doi={10.1109/TASLP.2021.3122291}}

@ARTICLE{9814838,
  author={Chen, Sanyuan and Wang, Chengyi and Zhengyang Chen et al.},
  journal={IEEE JSTSP}, 
  title={WavLM: Large-Scale Self-Supervised Pre-Training for Full Stack Speech Processing}, 
  year={2022},
  volume={16},
  number={6},
  pages={1505-1518},
  doi={10.1109/JSTSP.2022.3188113}}

@InProceedings{pmlr-v202-chen23ag,
  title = 	 {{BEAT}s: Audio Pre-Training with Acoustic Tokenizers},
  author =       {Chen, Sanyuan and Wu, Yu and Chengyi Wang et al.},
  booktitle = 	 {Proceedings of the 40th ICML},
  pages = 	 {5178--5193},
  year = 	 {2023},
  volume = 	 {202},
  series = 	 {Proceedings of Machine Learning Research},
  month = 	 {23--29 Jul},
  url = 	 {https://proceedings.mlr.press/v202/chen23ag.html}
}

@inproceedings{10.24963/ijcai.2024/421,
author = {Chen, Wenxi and Liang, Yuzhe and Ziyang Ma et al.},
title = {EAT: self-supervised pre-training with efficient audio transformer},
year = {2024},
isbn = {978-1-956792-04-1},
url = {https://doi.org/10.24963/ijcai.2024/421},
doi = {10.24963/ijcai.2024/421},
booktitle = {Proceedings of the Thirty-Third IJCAI},
articleno = {421},
numpages = {9},
location = {Jeju, Korea}
}

@article{WANG2026101825,
title = {ASVspoof 5: Design, collection and validation of resources for spoofing, deepfake, and adversarial attack detection using crowdsourced speech},
journal = {Computer Speech \& Language},
volume = {95},
pages = {101825},
year = {2026},
issn = {0885-2308},
doi = {https://doi.org/10.1016/j.csl.2025.101825},
url = {https://www.sciencedirect.com/science/article/pii/S0885230825000506},
author = {Xin Wang and Héctor Delgado and Hemlata Tak et al.}
}

@inproceedings{9123b6b9056f4f0291698dc291d0fc85,
title = "ADD 2023: the Second Audio Deepfake Detection Challenge",
author = "Jiangyan Yi and Jianhua Tao and Ruibo Fu et al.",
year = "2023",
pages = "125--130",
  booktitle={DADA@IJCAI}
}

@inproceedings{han2023,
  author={Soyul Han and Taein Kang and Sunmook Choi et al.},
  title={CAU KU Deepfake Detection System for ADD 2023 Challenge},
  year={2023},
  cdate={1672531200000},
  pages={23-30},
  url={https://ceur-ws.org/Vol-3597/paper5.pdf},
  booktitle={DADA@IJCAI}
}

@ARTICLE{10502279,
  author={Yang, Shu-wen and Chang, Heng-Jui and Zili Huang et al.},
  journal={IEEE/ACM TASLP}, 
  title={A Large-Scale Evaluation of Speech Foundation Models}, 
  year={2024},
  volume={32},
  number={},
  pages={2884-2899},
  doi={10.1109/TASLP.2024.3389631}}

@INPROCEEDINGS{10448453,
  author={Shin, Hyun-seo and Heo, Jungwoo and Ju-ho Kim et al.},
  booktitle={ICASSP 2024}, 
  title={HM-CONFORMER: A Conformer-Based Audio Deepfake Detection System with Hierarchical Pooling and Multi-Level Classification Token Aggregation Methods}, 
  year={2024},
  volume={},
  number={},
  pages={10581-10585},
  doi={10.1109/ICASSP48485.2024.10448453}}

@inproceedings{martindonas24_interspeech,
  title     = {{Exploring Self-supervised Embeddings and Synthetic Data Augmentation for Robust Audio Deepfake Detection}},
  author    = {Juan M. Martín-Doñas and Aitor Álvarez and Eros Rosello et al.},
  year      = {2024},
  booktitle = {{Interspeech 2024}},
  pages     = {2085--2089},
  doi       = {10.21437/Interspeech.2024-942},
  issn      = {2958-1796},
}

@inproceedings{wu24b_interspeech,
  title     = {{Spoofing Speech Detection by Modeling Local Spectro-Temporal and Long-term Dependency}},
  author    = {Haochen Wu and Wu Guo and Zhentao Zhang et al.},
  year      = {2024},
  booktitle = {{Interspeech 2024}},
  pages     = {507--511},
  doi       = {10.21437/Interspeech.2024-251},
  issn      = {2958-1796},
}

@inproceedings{Park_2019, 
   title={SpecAugment: A Simple Data Augmentation Method for Automatic Speech Recognition},
   url={http://dx.doi.org/10.21437/Interspeech.2019-2680},
   DOI={10.21437/interspeech.2019-2680},
   booktitle={Interspeech 2019},
   author={Park, Daniel S. and Chan, William and Zhang, Yu et al.},
   year={2019},
   month=sep, pages={2613–2617} }

@misc{jung2021aasistaudioantispoofingusing,
      title={AASIST: Audio Anti-Spoofing using Integrated Spectro-Temporal Graph Attention Networks}, 
      author={Jee-weon Jung and Hee-Soo Heo and Hemlata Tak et al.},
      year={2021},
      eprint={2110.01200},
      archivePrefix={arXiv},
      primaryClass={eess.AS},
      url={https://arxiv.org/abs/2110.01200}, 
}

@misc{kong2020hifigangenerativeadversarialnetworks,
      title={HiFi-GAN: Generative Adversarial Networks for Efficient and High Fidelity Speech Synthesis}, 
      author={Jungil Kong and Jaehyeon Kim and Jaekyoung Bae},
      year={2020},
      eprint={2010.05646},
      archivePrefix={arXiv},
      primaryClass={cs.SD},
      url={https://arxiv.org/abs/2010.05646}, 
}

@inproceedings{
lee2023bigvganuniversalneuralvocoder,
title={Big{VGAN}: A Universal Neural Vocoder with Large-Scale Training},
author={Sang-gil Lee and Wei Ping and Boris Ginsburg et al.},
booktitle={The Eleventh ICLR },
year={2023},
url={https://openreview.net/forum?id=iTtGCMDEzS_}
}

@inproceedings{jang2021univnetneuralvocodermultiresolution,
  title     = {UnivNet: A Neural Vocoder with Multi-Resolution Spectrogram Discriminators for High-Fidelity Waveform Generation},
  author    = {Won Jang and Dan Lim and Jaesam Yoon et al.},
  year      = {2021},
  booktitle = {Interspeech 2021},
  pages     = {2207--2211},
  doi       = {10.21437/Interspeech.2021-1016},
  issn      = {2958-1796},
}

@INPROCEEDINGS{10446331,
  author={Wang, Xin and Yamagishi, Junichi},
  booktitle={ICASSP 2024}, 
  title={Can Large-Scale Vocoded Spoofed Data Improve Speech Spoofing Countermeasure with a Self-Supervised Front End?}, 
  year={2024},
  volume={},
  number={},
  pages={10311-10315},
  doi={10.1109/ICASSP48485.2024.10446331}}

@INPROCEEDINGS{10094779,
  author={Wang, Xin and Yamagishi, Junichi},
  booktitle={ICASSP 2023}, 
  title={Spoofed Training Data for Speech Spoofing Countermeasure Can Be Efficiently Created Using Neural Vocoders}, 
  year={2023},
  volume={},
  number={},
  pages={1-5},
  doi={10.1109/ICASSP49357.2023.10094779}}

@misc{yin2025esdd2026environmentalsound,
      title={ESDD 2026: Environmental Sound Deepfake Detection Challenge Evaluation Plan}, 
      author={Han Yin and Yang Xiao and Rohan Kumar Das et al.},
      year={2025},
      eprint={2508.04529},
      archivePrefix={arXiv},
      primaryClass={cs.SD},
      url={https://arxiv.org/abs/2508.04529}, 
}

\end{document}